\documentclass[aps,prb,reprint,showpacs,amsmath,amssymb,superscriptaddress]{revtex4-1}
\usepackage{amsmath, mathrsfs, latexsym, amssymb, color, graphicx, subfigure, comment, hyperref, scrhack}
\usepackage{ifpdf}
\usepackage{physics}
\usepackage{float}

\begin{document}

\date{\today}

\title{Linear dynamics of classical spin as M\"{o}bius transformation}

\author{Alexey Galda}
\affiliation{James Franck Institute, University of Chicago, Chicago, Illinois 60637, USA}
\affiliation{Materials Science Division, Argonne National Laboratory, Argonne, Illinois 60439, USA}
\author{Valerii M. Vinokur}
\affiliation{Materials Science Division, Argonne National Laboratory, Argonne, Illinois 60439, USA}

\newcommand\ii{\mathrm{i}}
\newcommand\ee{\mathrm{e}}

\def\m{\mathbf{m}}
\def\M{\mathbf{M}}
\def\s{\mathbf{s}}
\def\S{\mathbf{S}}
\def\h{\mathbf{h}}
\def\H{\mathcal{H}}
\def\PT{\mathcal{PT}}
\def\P{\mathcal{P}}
\def\T{\mathcal{T}}
\def\C{\mathbb{C}}
\def\iis{\mathbf{i_\text{s}}}
\def\is{i_\text{s}}
\def\zz{\mathbf{\hat{z}}}
\def\Hext{\mathbf{H_\text{ext}}}
\def\Heff{\mathbf{H_\text{eff}}}
\def\HH{\mathbf{H}}
\def\heff{\mathbf{h_\text{eff}}}
\def\theff{\tilde{\mathbf{h}}_\text{eff}}
\def\Ms{M_\text{s}}
\def\ep{\mathbf{e_p}}
\def\ex{\mathbf{e_x}}
\def\ey{\mathbf{e_y}}
\def\ez{\mathbf{e_z}}
\def\j{\mathbf{j}}
\def\ep{\mathbf{e_p}}
\def\muB{\mu_\text{B}}

\begin{abstract}
Although the overwhelming majority of natural processes occurs far from the equilibrium, general theoretical approaches to non-equilibrium phase transitions remain scarce. Recent breakthroughs introducing description of open dissipative systems in terms of non-Hermitian quantum mechanics allowed to identify a class of non-equilibrium phase transitions associated with the loss of combined parity (reflection) and time-reversal symmetries.
Here we report that time evolution of a single classical spin (e.g. monodomain ferromagnet) governed by the Landau-Lifshitz-Gilbert-Slonczewski equation in absence of higher-order anisotropy terms is described by a M\"{o}bius transformation in complex stereographic coordinates. We identify the \textit{parity-time} symmetry-breaking phase transition occurring in spin-transfer torque-driven linear spin systems as a transition between hyperbolic and loxodromic classes of M\"{o}bius transformations, with the critical point of the transition corresponding to the parabolic transformation.
This establishes the understanding of non-equilibrium phase transitions as topological transitions in configuration space.

\end{abstract}

	  
\maketitle

The interest to dissipative spin-transfer torque (STT)-driven dynamics of a spin described by Landau-Lifshitz-Gilbert-Slonczewski (LLGS) equation~[\onlinecite{LL, Gilbert, Slon96}] is two-fold. On the applied side, the spin controlled by the applied spin-polarized current is an elemental unit for a wealth of spintronic applications. On the fundamental side, complete quantitative understanding of single-spin dynamics provides the essential tool for predictive description of a wealth of complex spin systems. Most notably, the nonconservative effect of Slonczewski STT on spin systems, equivalent to the action of \textit{imaginary} magnetic field, offers a unique tool for studying Lee-Yang zeros~[\onlinecite{Lee-Yang}] in ferromagnetic Ising and Heisenberg models.

It has recently been shown that nonequilibrium classical spin dynamics described by the LLGS equation naturally follows from the non-Hermitian extension of Hamiltonian formalism~[\onlinecite{GV16}]. Within this approach, the nonconservative effects of Gilbert damping and applied Slonczewski STT~[\onlinecite{Slon96}] originate from the imaginary part of system's Hamiltonian. This new technique has enabled important advances in the field of nonlinear spin dynamics, including the discovery of \textit{parity-time} ($\PT$) symmetry-breaking in systems with mutually orthogonal applied magnetic field and STT. This new type of phase transitions in spin systems is possible due to the fact that the action of STT is invariant under simultaneous operations of time-reversal and reflection with respect to the direction of spin polarization. Here we find that the $\PT$ symmetry-breaking phase transition occurring in STT-driven linear spin systems~[\onlinecite{GV16}] is a transition between hyperbolic and loxodromic classes of M\"{o}bius transformations governing the spin dynamics, with the critical point of the transition corresponding to the parabolic transformation. This establishes that non-equilibrium phase transitions associated with $\PT$ symmetry breaking are topological transitions in the configuration space.

We undertake the analytical study of dissipative STT-driven dynamics of a single classical spin described by a linear (in spin operators) non-Hermitian Hamiltonian. We show that in the absence of higher-order anisotropies, the combined effect of external magnetic field, Gilbert damping and applied Slonczewski STT can be incorporated in the effective action of a \textit{complex} magnetic field. In complex stereographic coordinates the equation of motion becomes a Riccati equation, which admits an exact solution in the form of a M\"{o}bius transformation of $\C^2$. The correspondence between various regimes of spin dynamics and classes of M\"{o}bius transformations is established and illustrated on the example of $\PT$ symmetry-breaking phenomenon, which is identified as a transition between elliptic and loxodromic M\"{o}bius transformations via a parabolic one.

The equation of motion can also be recast into the linear form by employing complex homogeneous coordinates without any approximations beyond the initial choice of the non-Hermitian spin Hamiltonian. The linear form of the spin dynamics equation provides a solid foundation for studying nonlinear effects in single and coupled spin systems (e.g. chaotic dynamics~[\onlinecite{Yang}, \onlinecite{Bragard}], spin-wave instabilities~[\onlinecite{Bertotti}], solitons~[\onlinecite{Lakshmanan}], etc.)

We study the most general \textit{linear} version of the spin Hamiltonian proposed in Ref.~[\onlinecite{GV16}],
\begin{equation}\label{H}
	\hat\H = \left( \frac{\gamma\HH + i\,\j}{1 - i\,\alpha}\right)\! \cdot \hat\S\,,
\end{equation}
where $\HH$ is the applied magnetic field, imaginary field $i\,\j$ is responsible for the action of STT, and phenomenological constant $\alpha$ describes Gilbert damping. The corresponding LLGS equation of spin dynamics reads
\begin{equation}\label{LLGS}
	\left(1 \!+\! \alpha^2\right)\!\dot{\S} \!=\! \gamma\,\HH \times \S \!-\! \frac{\alpha\gamma}{S}\,\S \times [\HH \times \S] \!+\! \frac1{S}\,\S \times [\S \times \j] \!+\! \alpha\,\S \times \j\,,
\end{equation}
where ${\gamma = g\muB/\hbar}$ is the absolute value of the gyromagnetic ratio, ${g \simeq 2}$, and $S \equiv |\S|$ is the total spin (constant in time). The first two terms in Eq.~(\ref{LLGS}) describe the standard Landau-Lifshitz (LL) torque and dissipation in Gilbert form, while the last two are responsible for Slonczewski STT, both dissipative (anti-damping) and conservative (effective field) contributions, respectively.

To show that Hamiltonian~(\ref{H}) yields the above LLGS dynamics equation in the classical limit (${S \to \infty}$), it is most convenient to consider $SU(2)$ spin-coherent states~[\onlinecite{Lieb}, \onlinecite{Stone}] ${\ket{\zeta} = \ee^{\zeta \hat S_+}\ket{S, -S}}$, where ${\hat S_\pm = \hat S_x \pm i\hat S_y}$, and ${\zeta \in \C}$ is the standard stereographic projection of the spin direction on a unit sphere, ${\zeta = (s_x + is_y)/(1 - s_z)}$, with the south pole (spin-down state) corresponding to ${\zeta = 0}$.

The Hamiltonian function in spin-coherent states reads
\begin{equation}
	\H (\zeta, \bar \zeta) = \frac{\expval{\hat\H}{\zeta}}{\bra{\zeta}\ket{\zeta}}\,,
\end{equation}
which yields~[\onlinecite{GV16}] the following compact form of Hamilton's equation of motion for classical spin:
\begin{equation}\label{zHamilton}
	\dot \zeta = i\frac{\left(1 + |\zeta|^2\right)^2}{2S} \frac{\partial \H}{\partial \bar \zeta}\,,
\end{equation}
where the factor $\left(1 + |\zeta|^2\right)^2\!/2S$ ensures invariance of measure on a two-sphere.

Let us now normalize and rewrite the linear non-Hermitian Hamiltonian~(\ref{H}) in terms of dimensionless variables:
\begin{equation}\label{H0}
	\hat\H_0 \equiv \hat\H/S = \widetilde \h \cdot \hat\s\,,
\end{equation}
where $\s \equiv \S/S$, and the effects of the applied magnetic field, Gilbert damping and Slonczewski STT contributions are all incorporated in the complex magnetic field, $\widetilde \h = (\tilde h_x, \tilde h_y, \tilde h_z) \in \C$.

The equation of motion~(\ref{zHamilton}) for the linear classical spin Hamiltonian~(\ref{H0}) can be rewritten as a linear matrix ordinary differential equation:
\begin{align}
	&\frac{d}{dt}\begin{bmatrix} \xi(t) \\ \eta(t) \end{bmatrix} = A \begin{bmatrix} \xi(t) \\ \eta(t) \end{bmatrix}\,,\label{homo}\\
	&A = \frac{i}{2}\sum_{k = x,y,z} \!\!\tilde h_k\,\sigma_k\,,\label{A}
\end{align}
where $\sigma_k$ are Pauli matrices and $\zeta(t) \equiv \xi(t)/\eta(t)$. The pair of complex functions $\{\xi, \eta\}$ are called homogeneous coordinates of $\zeta$~[\onlinecite{Needham}], such that each ordered pair $\{\xi, \eta\}$ (except $\{0, 0\}$) corresponds to a unique stereographic projection coordinate $\zeta$. The initial conditions for Eq.~(\ref{homo}) can be chosen as ${\xi(0) = \zeta(0),\,\eta(0) = 1}$.

The solution in terms of stereographic projection coordinates $\zeta$ has a simple form of a M\"{o}bius transformation:
\begin{equation}
	\zeta(t) = \frac{M_{11} \zeta(0) + M_{12}}{M_{21} \zeta(0) + M_{22}} \equiv M\!\left[\zeta(0)\right]\,,
\end{equation}
where the normalized ($\tr M = 1$) transformation matrix is given by the matrix exponential:
\begin{equation}\label{Mat}
	M = \ee^{At}\,.
\end{equation}

\begin{figure*}[!bht]
	\includegraphics[width=2\columnwidth]{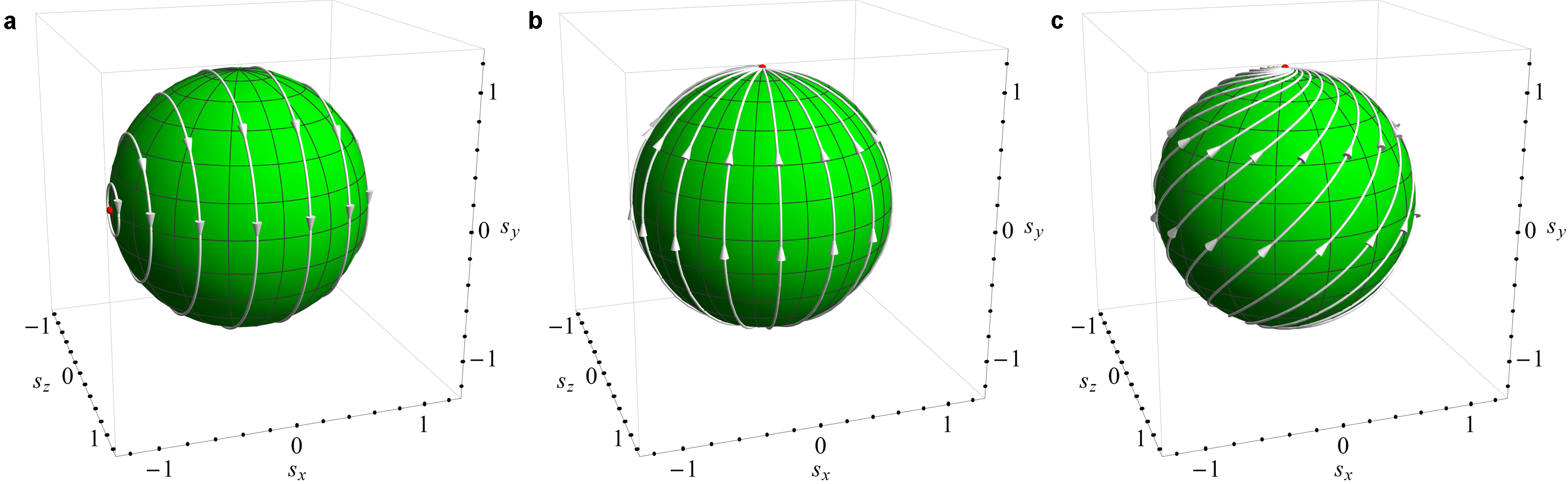}
	\caption{Geodesics of an elliptic (a), hyperbolic (b), and loxodromic (c) M\"{o}bius transformation, corresponding in spin dynamics to the applied real magnetic field along the $x$ axis (a), imaginary magnetic field along the $y$ axis (b), and complex magnetic field along the $y$ axis (c).}
	\label{fig1}
\end{figure*}

The equation of motion (\ref{homo}) illustrates that the discussed here classical spin dynamics can be written in a linear form, despite the nonlinear nature of the LLGS equation~(\ref{LLGS}) it reproduces. Understanding this linear system and its solutions presents a crucial step in describing nonlinear STT-driven magnetic systems.

\section*{M\"{o}bius transformation}

We now study the solution of Eq.~(\ref{zHamilton}) for linear spin Hamiltonians. Without loss of generality, we can take $\tilde h_z = 0$ and ${\mathrm{Im}\,(\tilde h_x) = 0}$ in Eq.~(\ref{H0}) by choosing the $z$ axis along $\bigl[\mathrm{Re}\,(\widetilde \h) \times \mathrm{Im}\,(\widetilde \h)\bigr]$ and $y$ axis along $\mathrm{Im}\,(\widetilde \h)$, while ${h_x \equiv \mathrm{Re}\,(\tilde h_x)}$ and ${\tilde h_y \in \C}$ can be arbitrary:
\begin{equation}
	\hat\H_0 = h_x \hat s_x + \tilde h_y  \hat s_y\,.\label{HH0}
\end{equation}

The equation of motion for this Hamiltonian takes the form of a Riccati equation:
\begin{equation}
	\dot \zeta(t) = -i\,\frac{h_x -i \tilde h_y}{2}\left[ \zeta^2(t) - \frac{h_x + i \tilde h_y}{h_x - i \tilde h_y}\right]\label{zdot}
\end{equation}
with two fixed points,
\begin{equation}\label{fixed}
	\zeta_{1,2} = \pm \sqrt{\frac{h_x + i \tilde h_y}{h_x - i \tilde h_y}}\,,
\end{equation}
and the solution
\begin{equation}\label{Mobius}
	\zeta(t) \!=\! \frac{\cos\!\left( \frac{\sqrt{h_x^2 + \tilde h_y^2}}{2}t\right)\zeta(0) \!+\! \frac{ih_x + \tilde h_y}{\sqrt{h_x^2 + \tilde h_y^2}}\sin\!\left( \frac{\sqrt{h_x^2 + \tilde h_y^2}}{2}t\right)}{\frac{ih_x - \tilde h_y}{\sqrt{h_x^2 + \tilde h_y^2}}\sin\!\left( \frac{\sqrt{h_x^2 + \tilde h_y^2}}{2}t\right) \zeta(0) \!+\! \cos\!\left( \frac{\sqrt{h_x^2 + \tilde h_y^2}}{2}t\right)}\,.
\end{equation}

As follows from Eq.~(\ref{Mobius}), the time evolution of a classical spin generated by an arbitrary linear non-Hermitian Hamiltonian is governed by a M\"{o}bius transformation of $\mathbb{C}^2$, when written in stereographic projection coordinates:
\begin{align}\label{M}
	[M] &= \left[ \ee^{\frac{i}{2}\left( h_x\sigma_x + \tilde h_y \sigma_y\right) t}\right]\notag\\
	&= \begin{bmatrix} \cos\!\left( \frac{\sqrt{h_x^2 + \tilde h_y^2}}{2}t\right) &  \frac{ih_x + \tilde h_y}{\sqrt{h_x^2 + \tilde h_y^2}}\sin\!\left( \frac{\sqrt{h_x^2 + \tilde h_y^2}}{2}t\right)\! \\
	\!\frac{ih_x - \tilde h_y}{\sqrt{h_x^2 + \tilde h_y^2}}\sin\!\left( \frac{\sqrt{h_x^2 + \tilde h_y^2}}{2}t\right) & \cos\!\left( \frac{\sqrt{h_x^2 + \tilde h_y^2}}{2}t\right) \end{bmatrix}\,,
\end{align}
in accordance with Eqs.~(\ref{A}), (\ref{Mat}), and~(\ref{HH0}).

\section*{Classification of M\"{o}bius transformations based on spin dynamics}

The traditional classification of M\"{o}bius transformations is based on the number and type of fixed points and distinguishes three different classes: elliptic, loxodromic (including hyperbolic as a special case) and parabolic transformations, which can be identified by calculating $\tr^2 M$~[\onlinecite{Needham}]. Here we show that all M\"{o}bius transformations can be obtained from a superposition of only two basic transformations, elliptic and hyperbolic, which in spin dynamics translate to applied real and imaginary magnetic fields, correspondingly. Elliptic M\"{o}bius transformation induces a uniform rotation of the entire Riemann sphere around a central axis, while a hyperbolic transformation produces antipodal expansion and contraction centers, see Fig.~\ref{fig1}, where the lines depict invariant geodesics of the corresponding M\"{o}bius transformation on the sphere. According to this consideration, every elliptic and hyperbolic transformation is fully determined by two parameters: `direction' and `amplitude', which define the direction of geodesics, including location of the fixed points, and the displacement of points on the Riemann sphere along geodesics upon the transformation.

In these terms, any M\"{o}bius transformation can be regarded as a superposition of only elliptic and hyperbolic transformations. A general loxodromic transformation has two fixed points, an attractive and repulsive nodes, and can be obtained from such a superposition in a unique way, provided the symmetry axes of the elliptic and hyperbolic transformations are not mutually orthogonal. Therefore, the transformation~(\ref{M}) is loxodromic when both of the following two conditions are met: a) $\beta \equiv \Im \tilde h_y \ne 0$ and b) $h_y \equiv \Re \tilde h_y \ne 0$ if $h_x \ne 0$, which follows from the requirement $\tr^2 M \in \C \backslash [0, 4]$~[\onlinecite{Needham}].

\begin{figure*}[!tbh]
	\includegraphics[width=2\columnwidth]{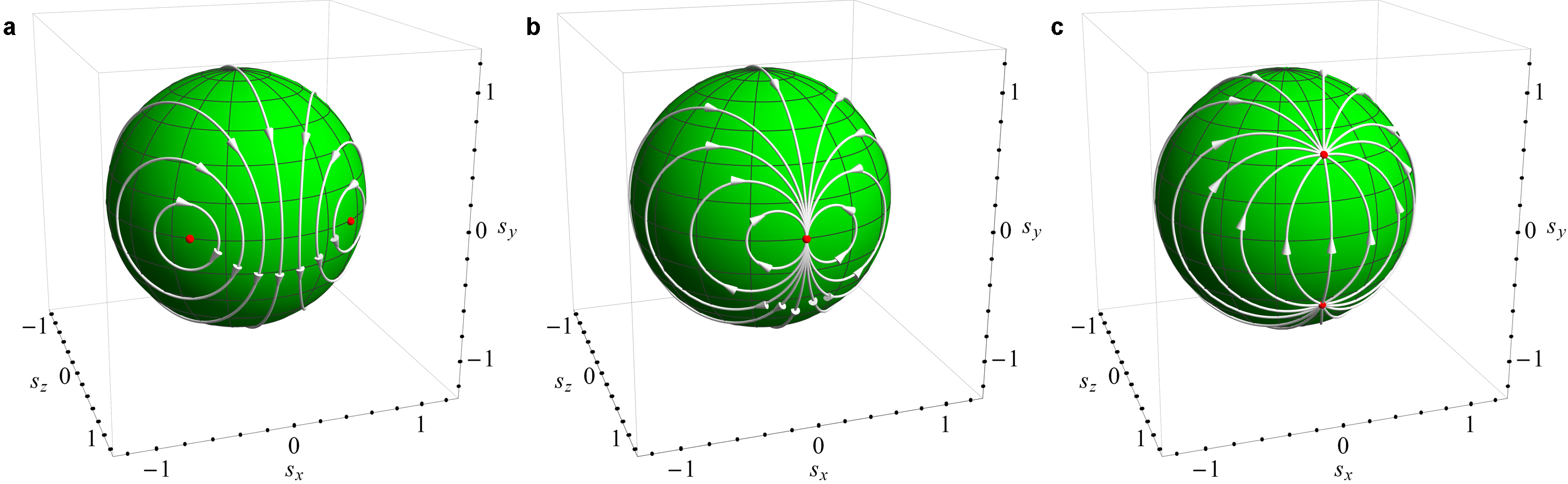}
	\caption{Transition between elliptic (a) and loxodromic (c) M\"{o}bius transformations by increasing $\epsilon$ past the critical value $1$, at which one obtains a parabolic transformation (b). In classical spin dynamics this transition corresponds to $\PT$ symmetry-breaking.}
	\label{fig2}
\end{figure*}

Let us now consider a superposition of mutually orthogonal elliptic and hyperbolic transformations. For the transformation~(\ref{M}) this corresponds to $h_x \neq 0$, $h_y = 0$, and $\beta \ne 0$. Depending on the ratio $\epsilon \equiv |\beta/h_x|$, the transformation (\ref{M}) can be elliptic ($\epsilon < 1$), loxodromic ($\epsilon > 1$) or parabolic ($\epsilon = 1$). As $\epsilon$ approaches 1 from below, two fixed points of the elliptic transformation move towards each other (see Fig.~\ref{fig2}a) until they eventually coalesce into a single fixed point of a parabolic transformation at $\epsilon = 1$, as shown in Fig.~\ref{fig2}b. As $\epsilon$ is increased further, the fixed point splits into an attractive and repulsive center of the hyperbolic transformation, see Fig.~\ref{fig2}c. In spin dynamics the described transition plays an important role. It is associated with the transition between regimes of unbroken and broken $\PT$ symmetry~[\onlinecite{GV16}].

Expectation values of the spin Hamiltonian~(\ref{H}) evaluated at the fixed points, Eq.~(\ref{fixed}), $E_{1,2} = \pm\sqrt{h_x^2 + \tilde h_y^2}$, are directly related to the eigenvalues of the corresponding M\"{o}bius transformation matrix, Eq.~(\ref{M}), ${\lambda_{1, 2} = \ee^{i\frac{E_{1,2}t}{2}}}$. They fully determine the type of the fixed points and that of the transformation. The standard classification~[\onlinecite{Needham}] uses multipliers of the transformation,
\begin{equation}
	\kappa_{1,2} \equiv \lambda_{1,2}^{-2} = \ee^{-i E_{1,2}t}\,,
\end{equation}
such that $|\kappa_{1,2}| = 1$ $\left( \kappa_{1,2} = \ee^{\pm i\theta} \neq 1\right)$ for elliptic transformations, $\kappa_{1,2} = 1$ for parabolic transformations, and $|\kappa_{1,2}| \neq 1$ for loxodromic transformations (with real $\kappa_{1,2} \neq 1$ in the special case of hyperbolic transformations). This fully agrees with the above considerations in the language of classical spin dynamics.

To conclude, we have shown that the time evolution of linear classical single-spin systems has a simple interpretation in terms of M\"{o}bius transformations of $\C^2$. The $\PT$ symmetry-breaking phase transition in such systems can be identified as a transition between elliptic and hyperbolic (via parabolic) classes of M\"{o}bius transformations appearing as solutions of the corresponding spin dynamics equations in complex stereographic coordinates. The established correspondence between linear spin dynamics and M\"{o}bius transformations reveals that any M\"{o}bius transformation can be produced by a unique superposition of an elliptic and hyperbolic transformations, corresponding to real and imaginary magnetic fields applied to a spin, respectively. We have demonstrated that the nonlinear LLGS equation describing dissipative STT-driven dynamics of a linear single-spin system can be written in a linear form, illustrating that such dynamics cannot produce any nonlinear effects, e.g. chaotic dynamic, for which additional time-dependent perturbation are necessary~[\onlinecite{Yang}, \onlinecite{Bragard}].

\section*{Acknowledgements}

This work was supported by the U.S. Department of Energy, Office of Science, Basic Energy Sciences, Materials Sciences and Engineering Division.

\makeatletter

\bibliographystyle{unsrt}

\end{document}